# Multiferroicity in the spin-1/2 quantum matter of LiCu$_2$O$_2$

**A. Rusydi, I. Mahns, S. Müller, and M. Rübhausen**

*Institut für Angewandte Physik, Universität Hamburg, Jungiusstraße 11, D-20355 Hamburg, Germany*

**S. Park, Y.J. Choi, C. L. Zhang, and S-W. Cheong**

*Department of Physics & Astronomy, Rutgers University, Piscataway, New Jersey 08854, USA*

**S. Smadici, P. Abbamonte**

*Physics Department and Frederick Seitz Materials Research Laboratory, University of Illinois, Urbana, IL, 6180, USA*

**M. v. Zimmermann**

*Hamburger Synchrotronstrahlungslabor HASYLAB at Deutsches Elektronen-Synchrotron DESY, Notkestraße 85, 22603 Hamburg, Germany*

**G. A. Sawatzky**

*Department of Physics and Astronomy, University of British Columbia, Vancouver, British Columbia V6T-1Z1, Canada*

**Multiferroicity in LiCu$_2$O$_2$ single crystals is studied using resonant soft x-ray magnetic scattering, hard x-ray diffraction, heat capacity, magnetic susceptibility, and electrical polarization. Two magnetic transitions are found at 24.6 K ($T_1$) and 23.2 K ($T_2$). Our data are consistent with a sinusoidal spin structure at $T_2<T<T_1$ and with a helicoidal spin structure at $T<T_2$ giving rise to ferroelectricity. Surprisingly, above $T_2$ the correlation lengths of the spin structures increase as the temperature increases with dramatic changes of ~42% along the *c*-axis. Our results demonstrate the interplay of the geometrical frustration and the electronic and magnetic polarizations.**

Low-dimensional spin (S)=1/2 systems have posed some of the most challenging scientific questions in solid state physics. The interplay between frustration and quantum spin fluctuations results in a rich phase diagram and unusual magnetic properties. In a model with quantum S=1/2 chains and competing nearest-neighbor interactions $J_1$ and next-nearest-neighbor interactions $J_2$, one expects, depending on $J_1/J_2$, a gapless collinear phase, a gapped disordered dimmer liquid phase, or a quasi-long-range ordered helicoidal spin structure.[1]

In **Fig. 1(a)**, $LiCu_2O_2$ consists of an equal number of $Cu^{1+}$ and $Cu^{2+}$. The magnetic $Cu^{2+}$ ions carry S=1/2 and are located at the center of edge-sharing $CuO_4$ plaquettes and form two frustrated quasi-one dimensional (1D) S=1/2 chains along the *b*-axis. These chains are separated by $Li^{1+}$ ions to form double layers parallel to the *ab*-plane which are separated by $Cu^{1+}$ sites. Within each chain, the Cu-O-Cu angle is about 94°. As a result $J_2$ is weaker than $J_1$. The strength of the interaction between chains ($J_{DC}$) is not clear, however we expect $J_1$ to be ferromagnetic and $J_2$ to be antiferromagnetic based on analysis of the magnetic susceptibility $\chi(T)$.[2] This leads to frustration and favors helimagnetism.[1] A similar scenario was recently proposed for a very similar isostructural $NaCu_2O_2$.[3]

$LiCu_2O_2$ exhibits striking properties such as the presence of a spin-singlet liquid state[4], incommensurate (IC) magnetic order[5] as well as ferroelectricity.[6] However, all these properties are not well-understood. One problem is the intrinsic chemical disorder. Electron-spin resonance (ESR)[4] has shown the presence of a spin-singlet state with a spin gap of 6 meV at 23 K. Specific heat[4] and nuclear magnetic resonance (NMR)[7] show phase transitions at 24.2 K, 22.5 K, and 9 K. Recent neutron diffraction[5] found one transition to an incommensurate (IC) magnetic superstructure below 22 K with a propagation vector $Q=([2n+1]/2, k+\delta, l)$ where *n*, *k*, and *l* are integer and $\delta \sim 0.174$. Thus it is concluded that the magnetic superstructure is a helical in the *ab*-plane. This is inconsistent with the observation of ferroelectricity with an electric polarization along the *c*-direction below ~23 K.[6]

In an attempt to understand the coupling between lattice, charge, and spin degrees of freedom in S=1/2 quantum systems, we have studied a $LiCu_2O_2$ single crystal using polarization-dependent resonant soft x-ray magnetic scattering (RSXMS), hard x-ray diffraction (HXD), heat capacity, magnetic susceptibility, and electrical polarization. All experiments were done on the same sample.

The $LiCu_2O_2$ single crystal was grown by the self-flux method.[6] The HXD was done at the BW5 beamline of HASYLAB (DESY) with a photon energy of 100.5 keV. The lattice parameters of the orthorhombic structure at 10 K are *a*=5.6963, *b*=2.8497, and *c*=12.417 Å. The HXD highlights the high crystalline perfection and confirms the absence of crystalline impurity phases. The crystal is found to be microscopically twinned along the [1,1,0] plane with $a \sim 2b$. Below 23 K ($T_{FE}$), $LiCu_2O_2$ becomes ferroelectric (FE) and shows a small anomaly in the dielectric constant $\varepsilon$.[6] **Figure 2(b)** shows a changing dielectric polarization (*P*) along the *c* direction as function of temperature. However, the observed polarization (~4μC/m$^2$) is smaller by 2-3 orders of magnitude as compared to RMnO$_3$[8,9] and RMn$_2$O$_5$.[9,10]

Two magnetic transitions are found in the magnetic susceptibility, $\chi$ (**Fig. 2(a)**). For an applied magnetic field $H$ parallel to the $c$-axis ($H\|c$), the curve $d\chi_c/dT$ presents transitions, at 23.2 K ($T_2$) and 24.6 K ($T_1$), while for $H\|b$, the $d\chi_c/dT$ shows only a sharp transition at 23 K. The $T_2$ coincides with $T_{FE}$ from heat capacity measurements (**Fig. 2(a)**). Our sample does not shown any other transition below 23 K.[11]

RSXMS was done on a surface which was cleaved *in-situ* with (2,1,0) orientation at the beamline X1B of the National Synchrotron Light Source (Brookhaven) using a 10-axis, ultrahigh-vacuum-compatible diffractometer.[12,13] X-ray absorption spectra (XAS) were measured in-situ in the fluorescence yield mode at the Cu$L_{3,2}$ edges. We denote the reciprocal space with Miller indices ($H,K,L$), which represent a momentum transfer $\boldsymbol{Q} = (2\pi H/a, 2\pi K/b, 2\pi L/c)$. The angle of incoming ($\theta_{in}$) and outgoing photons ($\theta_{out}$) depends on $\boldsymbol{Q}$ but was approximately 35$^o$ and 55$^o$, respectively. The azimuthal angle, $\phi$, is $\phi=0^o$ and 90$^o$ (**Fig. 1(b)**).

Scattering at transition metal $L$ edges is known to be sensitive to the spin modulation.[14,15,16,17] **Figure** 1 (**c**) illustrates Cu $2p\rightarrow3d$ resonant scattering process which enhances the magnetic scattering from Cu$^{2+}$. In the cuprate systems, the Cu $2p\rightarrow3d$ transition exhibits two main peaks corresponding to final states with $2p_{3/2}$ and $2p_{1/2}$ core holes, referred to as the Cu$L_3$ and Cu$L_2$ absorption edges, respectively. This material is particularly interesting because it has a clear contrast in the scattering of the Cu$^{1+}$ and Cu$^{2+}$ sites (see **Fig. 2(d)**). The peaks at 930 eV and 950 eV are Cu$L_{3,2}$ edges of Cu$^{2+}$ sites and the peaks at 933 eV and 953 eV are Cu$L_{3,2}$ edges of Cu$^{1+}$ sites.[18]

Probing with a photon energy $E$ = 930 eV, an IC superstructure with Q = (0.5, 0.1738, 0) at T = 18 K is observed (**Fig. 2(c)**). This is identical to the magnetic superstructure found by neutron diffraction.[5] Our experiment reveals that the correlation lengths along $a$, $b$, and $c$ are very large with $\xi_a$=(1662±20), $\xi_b$=(2120±20), and $\xi_c$=(935±20) Å, respectively. X-rays at 930 eV have penetration depth of 2500 Å.

**Figure 2(d)** shows the scattering intensity of the superstructure ($I_{ss}$) as a function of photon energy, i.e. the resonance profile (RP), at 18 and 24.6 K across the Cu$L_3$ edge. The 24.6 K and 18 K measurements show magnetic scattering above and below the *FE* transition, respectively. The RP is compared to the complex atomic scattering factor of Cu$^{2+}$, $f_{Cu}(E)$. In this case, $I_{SS} \propto |f_{Cu}|^2 = |\text{Re}[f_{Cu}]+\text{Im}[f_{Cu}]|^2$. The Im[$f_{Cu}(E)$] is determined from the absorptive part of the refractive index, Im[$n$], which is linearly related to the XAS spectrum, through the relation $\text{Im}[n(E)] = -(r_e\lambda^2 N/2\pi V_{cell})\text{Im}[\sum_i f_i(E)]$. The Re[$f_{Cu}(E)$] is calculated from Im[$f_{Cu}(E)$] by performing a Kramers-Kronig transform. Here $r_e$ is the classical electron radius, $\lambda$ is the x-ray wavelength, $N$ is the number of Cu in the unit cell, and $f_i$ is the complex atomic scattering factor. The XAS measurement was done with an incident x-ray polarization in the *ab*-plane, at room temperature, corrected for self absorption, and placed on an absolute scale.[19] It is clear that the superstructure is greatly enhanced at the Cu$^{2+}$ peak. A gigantic enhancement occurs in the

magnetic-*FE* state. In a soft x-ray measurement a superstructure due to a lattice distortion would result in orders of magnitude smaller scattering intensities.[12]

We have performed a HXD study to rule out a lattice modulation.[20] Even at 4 K, neither a $(0.5, k\pm\delta, 0)$ nor a $(0.5, k\pm 2\delta, 0)$ reflection was found supporting that the lattice distortion is extremely small. This is in contrast to ferroelectric TbMnO$_3$ in which lattice distortions are observed.[21] This further supports the conclusion that the lattice related effects are very weak.

**Figure 3(a)** and **3(b)** display the intensity together with position of the Bragg peak as a function of temperature and the polarization of the incoming photon. For $\phi=0°$, we have found the presence of two magnetic transitions: at ~23.2 K and ~24.6 K which is consistent with the magnetic susceptibility. The intensity increases as the temperature decreases indicating an enhanced magnetic order upon cooling, while $Q$ also changes with temperature. For $\phi=90°$, the superstructure vanishes rapidly above $T_2$.

The combination of polarization-dependent RSXMS, HXD, magnetic susceptibility, and the electrical polarization measurements provides crucial information regarding the coexistence of *FE* and magnetic states. At $T_2<T<T_1$, we find: First, $d\chi/dT$ shows an anomaly at $T_1$ for $H\|c$ implying that the *c*-direction is an *easy* axis (**Fig. 2(a)**); Second, there is no *FE*, i.e. $P=0$ (**Fig. 2(b)**); Third, for $\phi=0°$ RSXMS experiment shows a magnetic Bragg reflection while for $\phi=90°$ the superstructures is very weak. For spins which have a component of the magnetic moment along the *c*-axis, the polarization factor of the magnetic scattering $f_{mag} = (\vec{e}_{in} \times \vec{e}_{out})\cdot\widehat{\mathbf{M}} \neq 0$. This implies that the spins have a strong component or are oriented along the *c*-axis; Fourth, no harmonics and no lattice distortion were found at $Q=2\delta$. A helical structure in *ac*-plane would result in a polarization independence. Therefore, we propose that the spin structure is a sinusoidal with the spin oriented along the *c*-axis and propagating with $Q$ along the *b*-direction.

For $T<T_2$ we find: First, $d\chi/dT$ shows the magnetic transitions for $H\|b$ and $H\|c$ implying that *b* and *c* are both the easy axis (**Fig. 2(b)**); Second, *FE* is present with $P_c\approx 3\mu C/m^2$ (**Fig. 3(a)**); Third, $f_{mag}$ is non-zero for $\phi=0°$ and $90°$, giving rise to a magnetic Bragg reflection; Fourth, no harmonics and no lattice distortion are found at $Q=2\delta$. These results cannot be explained with a simple spiral spin structure in the *bc*-plane as such a condition would result in $f_{mag}=0$ for $\phi=90°$. A pure helical or sinusoidal spin wave cannot cause ferroelectricity since $P\cdot\nabla M$ is zero.[22] Therefore, our results are consistent with a spin structure that is helicoidal having *a*, *b*, and *c* components, and propagating with Q along the *b* direction. The complexity of the helicoidal structure relates to a coupling of the two frustrated quasi-1D S=1/2 chains along the *b*-axis, as shown in **Fig. 1(a)**, resulting to a total polarization *P* along the *c*-axis. This highlights the underlying frustration. Frustration can be lifted by breaking the symmetry yielding an electronically driven ferroelectric order.

Moreover, we find a decreased correlation length in the *FE* state with decreasing temperature [see **Fig. 3(c)**-(**e**)]. A dramatic change of the coherence lengths occurs around $T_2$ which is the transition from sinusoidal to helicoidal spin structures. The strongest change is ~42% along *c*, followed by ~22% along *b*, and ~4% along the *a* directions. The strongest distortion is along the *c* direction which is also the direction of the ferroelectric moment. This is a surprising result as it cannot be explained by enhanced thermal fluctuations close to the phase transition. Usually, the correlation lengths are getting shorter at high temperatures due to fluctuations which eventually destroy the long range order. In our case, *FE* domains form below $T_2$ in the helicoidal spin phase disturbing the magnetic coherence lengths while the sinusoidal spin structure above $T_2$ is free of *FE* domains leading to an increased coherence length with increasing temperature.

Another interesting observation is the smallness of the temperature *window* between the two transitions. In TbMnO$_3$, S=2,[20] the temperature window is about 12 K. In the spinel CoCr$_2$O$_4$, S=3 [23] the temperature window is ~65K. This shows that the lack of a single ion magnetic annisotropy in S=1/2 systems results in an unstable sinusoidal phase supporting our picture of an electronically driven phase. Related to this is the increased coherence lengths with increasing temperature above $T_{FE}$ indicating remaining dynamic FE domain ordering.

In conclusion, we have found two magnetic transitions in LiCu$_2$O$_2$: a sinusoidal spin structure at $T_2<T<T_1$ and a helicoidal spin structure at $T<T_2$ giving rise to ferroelectricity. The coherence lengths of the superstructure are increasing as the temperature increases. We understand these phenomena by considering the interplay between magnetic and electronic ordering.

We acknowledge M. Mostovoy for helpful discussions and financial support by the Helmholtz Association VH-FZ-007, DFG Ru 773/2-3, DOE No. DE-FG02-06ER46285, NSLS No. DE-AC02-98CH10886, NSF-MDR-0405682, NSERC, CIAR, and CFI.

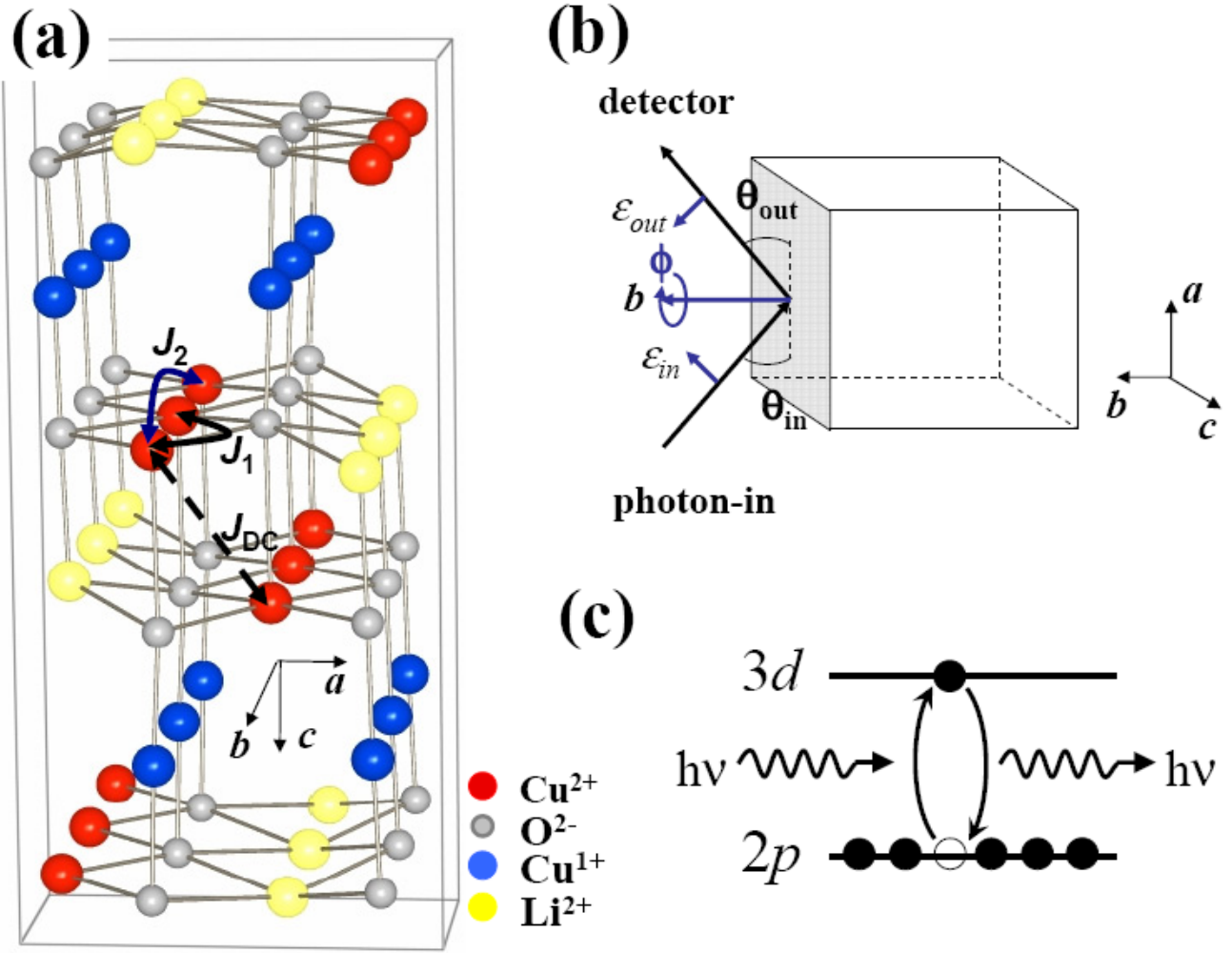

**Figure 1**. (**a**) The crystal structure of LiCu$_2$O$_2$ and the double chains showing $J_1$, $J_2$ and $J_{DC}$. (**b**) The scattering experimental geometry with photon polarization $\varepsilon$ in the scattering plane. The azimuthal angle, $\phi$, is $0^o$ in the geometry shown, where the photon polarization is perpendicular to $c$-axis.(**c**) An ilustration of Cu $2p \rightarrow 3d$ resonant scattering process.

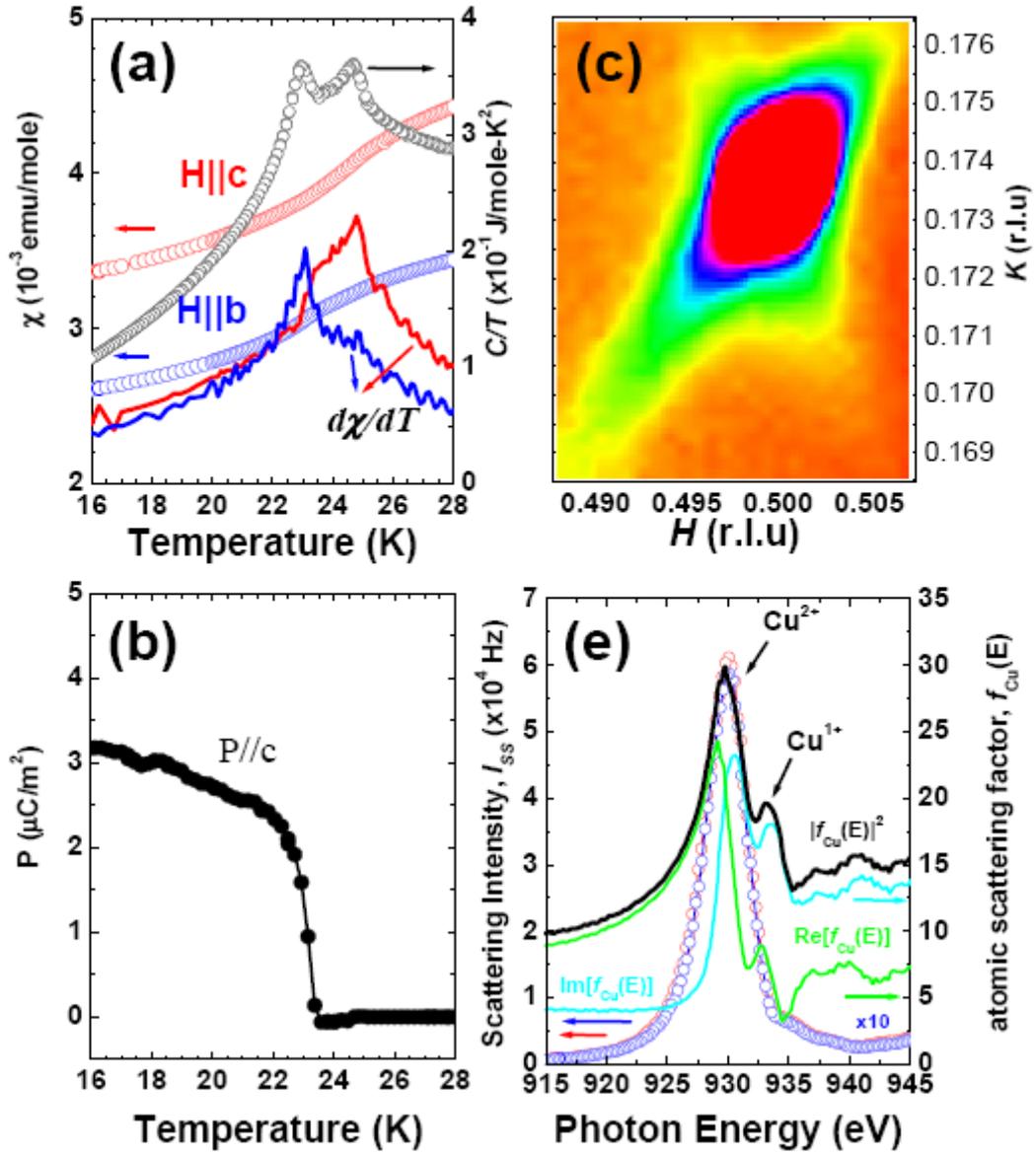

**Figure 2**. (**a**) magnetic susceptibility and heat capacity measurements showing two magnetic transitions, at 23.2 K and 24.6 K. (**b**) The electrical polarization changes along *c*-axis showing the onset of polarization at 23.2 K. (**c**) Appearance of the magnetic structure on resonance (E=930 eV ) in the reciprocal space map (H,K,L) = (0.5, 0.174,0) taken at 18 K. The correlation length along *n* direction, $\xi_n$ is defined as $n_n/\Delta q_n$, where $\Delta q_n$ is the width of the Bragg reflection and $n_n$ is the lattice parameter along n direction. (**d**) Energy scan at fixed-$Q$ = (0.5,δ,0) for two different temperatures: 18 K at δ=0.1738 (red dots) and 24.6 K at δ=0.1722 (blue dots) showing the same profile which indicates the lack of lattice distortion in the *FE* phase (T$_{FE}$ = 23.2 K) at this particular and direction of modulation. The energy dependence of the resonantly diffracted intensity is compared to the complex atomic scattering factor of Cu, $f_{Cu}(E)$, to highlight the resonant the $Cu^{2+}$ peak.

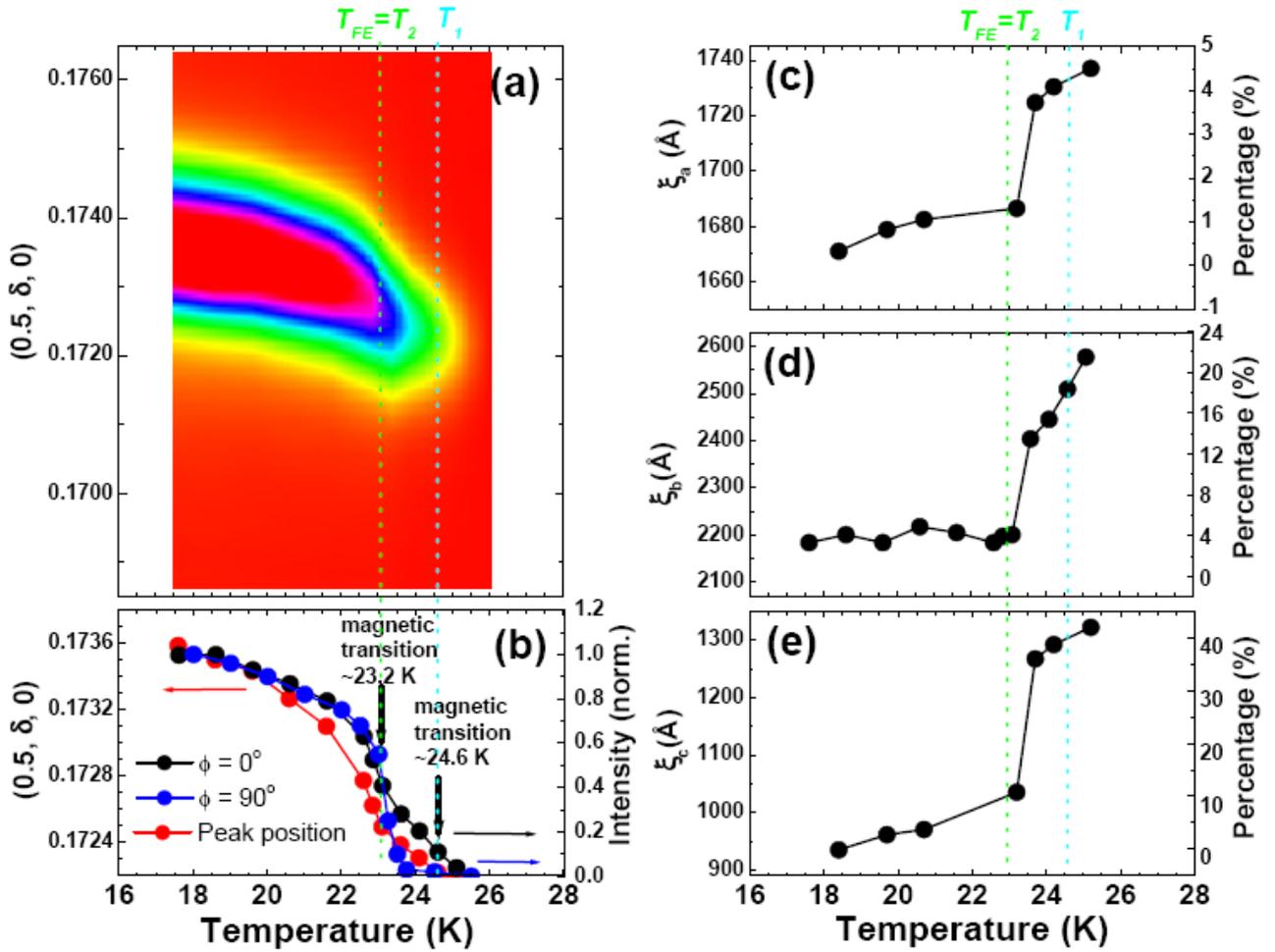

**Figure 3**. (**a**) Two-dimensional plot of temperature vs (0.5,*K*,0) for $\phi=90°$ showing the changing of peak position as function of temperature. (**b**) The evolution of the (red dots) peak position, $\delta$ and (black and blue dots) intensity of the magnetic scattering for the two polarizations as function of temperature showing two transitions: ~23.2 K and ~24.6 K. (**c**)-(**e**) Correlation lengths of the magnetic ordering at the transition into the FE state for all crystallographic directions.